\documentclass[%
 twocolumn ,
 amsmath,amssymb,
 aps,
prb,
]{revtex4-2}
\usepackage{tikz}

\usepackage{graphicx}
\usepackage{dcolumn}
\usepackage{bm}
\usepackage{comment}

\begin{document}

\preprint{APS/123-QED}

\title{Kolmogorov--Arnold networks in molecular dynamics
}

\author{Yuki Nagai}
\affiliation{Information Technology Center, The University of Tokyo, 6--2--3 Kashiwanoha, Kashiwa, Chiba 277--0882, Japan
}
\email{nagai.yuki@mail.u-tokyo.ac.jp}
\author{Masahiko Okumura}%
\affiliation{%
Center for Computational Science and e-Systems, Japan Atomic Energy Agency, Kashiwa, Chiba 277--0871, Japan
}%

\date{\today}

\begin{abstract}
We explore the integration of Kolmogorov--Arnold Networks (KANs) into molecular dynamics (MD) simulations to improve interatomic potentials. 
We propose that widely used potentials, such as the Lennard--Jones (LJ) potential, the embedded atom model (EAM), and artificial neural network (ANN) potentials, can be interpreted within the KAN framework. 
Specifically, we demonstrate that the descriptors for ANN potentials, typically constructed using polynomials, can be redefined using KAN’s non-linear functions. 
By employing linear or cubic spline interpolations for these KAN functions, 
we show that the computational cost of evaluating ANN potentials and their derivatives is reduced.
\end{abstract}

\maketitle
\section{Introduction}
Molecular dynamics (MD) simulations have revolutionized the field of materials science, providing detailed atomistic insights into the structure, properties, and behavior of materials under various conditions \cite{Rahman1971-be, Wang2010-ia, Kubicki1988-nx, Engel2015-ku, Kobayashi2023-fa}. 
By solving Newton's equations of motion for a system of interacting atoms, MD simulations can model processes such as diffusion, phase transitions, mechanical deformation, and chemical reactions. 
Despite their widespread utility, traditional MD simulations are often limited by the accuracy and efficiency of the force fields used to approximate interatomic forces.

{\it Ab initio} molecular dynamics (AIMD), particularly those based on density functional theory (DFT), offer enhanced accuracy by explicitly accounting for the electronic structure of materials \cite{Iftimie2005-zo, Mouvet2022-nm}. 
However, the computational cost of AIMD is prohibitive for simulating large systems or long-time scales, restricting its application to relatively small and short simulations. 
This limitation underscores the need for more efficient computational methods that do not compromise on accuracy.

Machine learning (ML) has emerged as a transformative approach to address these challenges in MD simulations. 
By training on energies obtained from DFT calculations, ML models, particularly artificial neural networks (ANNs), can accurately predict potential energy surfaces and interatomic forces \cite{Behler2007-kn}.
Integrating ML with MD, known as machine learning molecular dynamics (MLMD), offers a promising pathway to enhance both the accuracy and efficiency of simulations \cite{Nagai2020-gx, Nagai2024-ff, Artrith2017-ow, Batzner2022-ns, Musaelian2023-av, Bartok2010-po}.

Developing novel and effective ML methods tailored specifically for MD simulations is crucial to further unlocking the potential of MLMD \cite{Bartok2010-po, Nagai2020-gx, Batzner2022-ns, Musaelian2023-av, Tholke2022-bg, Sivaraman2020-nq, Wang2024-ey}.
For MLMD, multi-layer perceptrons (MLPs) are mainly used for ANN. 
Recently, Kolmogorov-Arnold networks (KANs), an alternative to MLPs, have been proposed \cite{Liu2024-ut}.
They are extensions of Kolmogorov--Arnold representation theorem \cite{Kolmogorov, Braun2009-pa}, which states that any continuous function can be represented as a superposition of continuous functions of a single variable.
This theoretical framework provides a powerful tool for decomposing complex functions into simpler, more manageable component functions.

In this paper, we explore the relations between KANs and potential energy in molecular dynamics. 
From the KAN perspective, the Lennard--Jones (LJ) potential, embedded atom mode (EAM), and ANN potential can all be viewed as forms of KANs. Utilizing linear or cubic spline interpolations, we can reduce the computational cost of the ANN potential. 

This paper is organized as follows. 
In Sec.~II, we introduce KANs. 
In Sec.~~III, we discuss the relation between the KANs and potentials used in molecular dynamics. 
The KAN descriptor is introduced in Sec.~~IV. 
The numerical results are shown in Sec.~V. We introduce the linear and cubic spline interpolations to approximate the KAN descriptor for the speedup of simulations. 
In Sec.~VI, the conclusion is given.

\section{Kolmogorov--Arnold Networks}
\subsection{Kolmogorov--Arnold representation theorem}
\begin{figure}
    \centering
    \includegraphics[width=0.75 \linewidth]{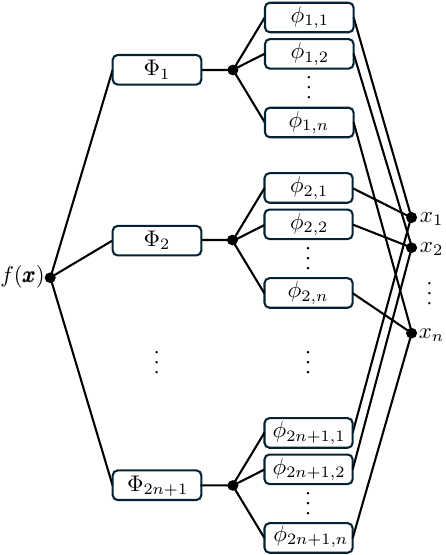}
    \caption{The schematic of Kolmogorov--Arnoold representation theorem.}
    \label{KART}
\end{figure}

Recently, KANs have been proposed inspired by the Kolmogorov--Arnold representation theorem \cite{Liu2024-ut}. 
The Kolmogorov--Arnold representation theorem shows that a multivariate function $f(\pmb{x}) = f(x_1, x_2, \ldots, x_n)$ can be represented by a superposition of univariate functions \cite{Kolmogorov, Braun2009-pa}.
For a smooth $f: [0,1]^n \rightarrow \mathbb{R}$, a function $f(\pmb{x})$ is represented as 
\begin{align}
    f(\pmb{x}) = \sum_{q=1}^{2n+1} \Phi_q \! \left( 
    \sum_{p=1}^{n} \phi_{q,p}(x_p)
    \right), \label{eq:kar}
\end{align}
where $\phi_{q,p}: [0,1] \rightarrow \mathbb{R}$, and $\Phi_q: \mathbb{R} \rightarrow \mathbb{R}$ (See, Fig.~\ref{KART}).
For example, one can easily confirm that the multiplication $x_1 x_2$ is represented by a superposition of univariate functions like Eq.~(\ref{KART}), 
\begin{align}
    x_1 x_2 &= \exp \! \left( \sum_{i=1}^2 \log x_i  \right).
\end{align}
\subsection{Kolmogorov--Arnold Networks}
\begin{figure*}[t]
\includegraphics[scale=0.7]{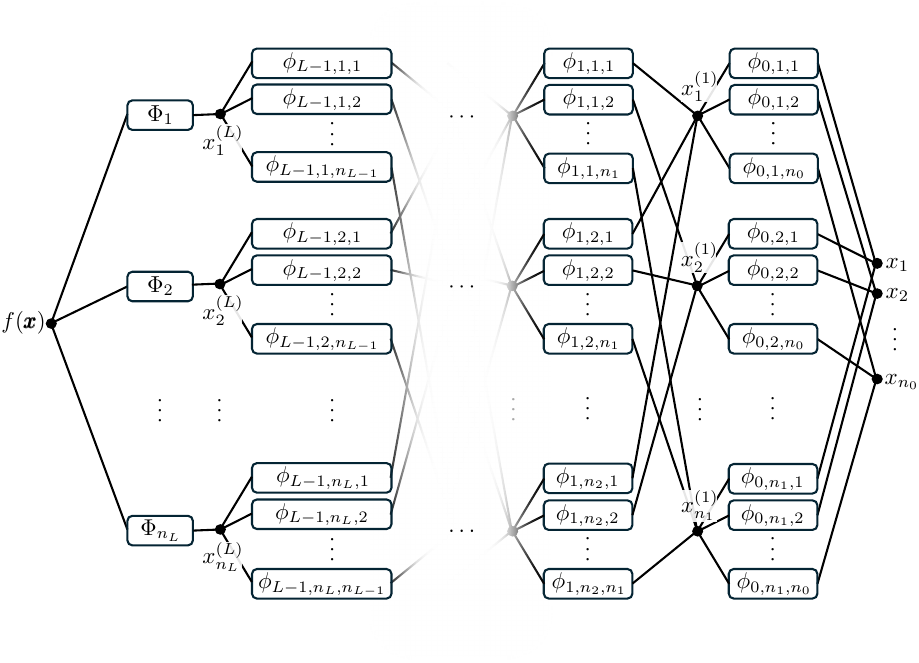}
\caption{The schematic of Kolmogorov--Arnold networks.}
\label{KAN_fig}
\end{figure*}

Liu {\it et al.} have extended the idea of the Kolmogorov--Arnold representation theorem \cite{Liu2024-ut}.
By regarding Eq.~(\ref{eq:kar}) as a two-layer network with one hidden layer with $2n + 1$ neurons, the more general form of the networks can be given by 
\begin{align}
    f({\bm x}) &= \sum_{i_L = 1}^{n_L} \Phi_{i_L} \! \left( x^{(L)}_{i_L} \right), \label{eq:phiL}\\
    x^{(L)}_{i_L} &= \sum_{i_{L-1}=1}^{n_{L-1}} \phi_{L-1,i_L,i_{L-1}} \! \left( x_{i_{L-1}}^{(L-1)} \right), \\
    x^{(L-1)}_{i_{L-1}} &= \sum_{i_{L-2}=1}^{n_{L-2}} \phi_{L-2,i_{L-1},i_{L-2}} \! \left( x_{i_{L-2}}^{(L-2)} \right), \\
    &\vdots \\
    x^{(2)}_{i_2} &= \sum_{i_{1}=1}^{n_{1}} \phi_{1,i_{2},i_{1}} \! \left( x_{i_{1}}^{(1)} \right),\\
    x^{(1)}_{i_1} &= \sum_{i_{0}=1}^{n_{0}} \phi_{0,i_{1},i_{0}}(x_{i_{0}}) \, , \label{eq:KANlayer}
\end{align}
which is called KAN.
The schematic of KAN is shown in Fig.~\ref{KAN_fig}.

The multivariate function $f({\bm x})$ is represented by a superposition of non-linear ``activation'' functions $\phi_{l,qp}(x)$. 
To construct function $f({\bm x})$, a shape of the non-linear functions $\phi_{l,qp}(x)$ is optimized. 
For example, in the implementation by Liu {\it et al}., each KAN layer is composed of the sum of the spline function and the SiLU activation function. 
There are various kinds of basis functions, such as Legendre polynomials, Chebyshev polynomials, and Gaussian radial distribution functions \cite{torchkan, fluxkan, Ss2024-su, Li2024-fi}.

\section{Potential energy in molecular dynamics}
\subsection{Various kinds of potential energies}
In the field of molecular dynamics, accurately estimating the forces on each atom is crucial. Since the force on the $i$-th atom at $\pmb{R}_i$ is derived from the derivative of the total potential energy $E$, i.e.,
\begin{equation}
\pmb{F}_i = - \frac{\partial E}{\partial \pmb{R}_i} \, .
\end{equation}
An accurate and fast estimation of the potential energy is one of the most important challenges.

Various kinds of potentials have been proposed for molecular dynamics. 
For example, the LJ potential is a simplified model that describes the essential features of interactions between simple atoms and molecules \cite{Schwerdtfeger2024-zg}. 
The EAM is an approximation describing the interatomic potential and is particularly appropriate for metallic systems \cite{Daw1984-mb}.

Calculations based on DFT, known as {\it ab initio} calculations, are among the most accurate methods for evaluating potential energy. AIMD has many applications in material design. However, reducing the computational effort required for AIMD remains a key issue for its broader application to phenomena on large lengths and time scales.

The use of ANNs, which imitate DFT energies through machine learning, is seen as a promising solution to this issue \cite{Behler2007-kn}. 
Research in machine-learning molecular dynamics has grown rapidly over the last decade \cite{Bartok2010-po, Nagai2020-gx, Batzner2022-ns, Musaelian2023-av, Tholke2022-bg, Sivaraman2020-nq, Wang2024-ey, Artrith2017-ow}. 
Various network structures have been proposed, including graph neural networks \cite{Takamoto2022-rq, Batatia2022-tb, Batzner2022-ns, Musaelian2023-av, Nagai2020-gx, Nagai2024-ff, Artrith2017-ow, Batzner2022-ns, Musaelian2023-av, Bartok2010-po} and transformer architectures \cite{Tholke2022-bg}.

\subsection{Relation between Kolmogorov--Arnold networks and potential energy}
\subsubsection{Lennard--Jones potential}
\begin{figure}
    \centering
    \includegraphics[width=0.5 \linewidth]{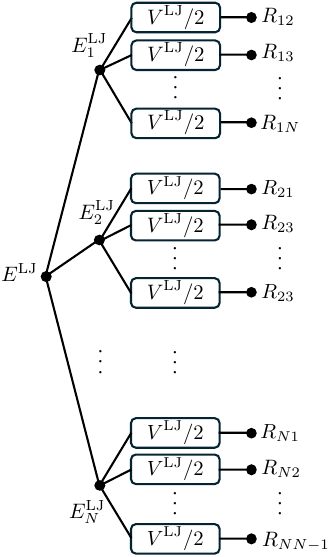}
    \caption{The schematic of Lennard--Jones potential.}
    \label{LJ}
\end{figure}

Let us discuss the relation between KAN and potential energy in molecular dynamics. Using the LJ potential, the total potential energy for $N$ atoms is expressed as
\begin{align}
    E^\mathrm{LJ}(\pmb{R}) &= \sum_{i=1}^{N} E_i^\mathrm{LJ} (\pmb{R}) \, ,\\
    E_i^\mathrm{LJ} (\pmb{R})  & = \frac{1}{2} \sum_{j=1, j \ne i}^N V_\mathrm{LJ} (R_{ij}) \, , 
\end{align}
where, with empirically determined coefficients $A_m$ and $B_n$, 
\begin{align}
V_\mathrm{LJ}(R) & = \frac{A_m}{R^m} - \frac{B_n}{R^n} \, ,\\
    R_{ij} & = |\pmb{R}_{ij}| \, , \\
    \pmb{R}_{ij} & = \pmb{R}_j - \pmb{R}_i \, .
\end{align}
The LJ potential can be represented by a type of KAN, as shown in Fig.~\ref{LJ}.

As shown in the KAN shown in Eq.~(\ref{eq:phiL}), a non-linear function depends on indices of both input and output layers $i_{L-1}$ and $i_{L}$. 
In the LJ potential model, the elements of the initial input are distances between two atoms, which should have a permutation symmetry. 
Therefore, all non-linear functions are $V_\mathrm{LJ}(R)$, similar to weight sharing in MLP networks. 
We call this feature ``function sharing.''

\subsubsection{Embedded atom model potential}
\begin{figure}[t]
    \centering
    \includegraphics[width=0.7 \linewidth]{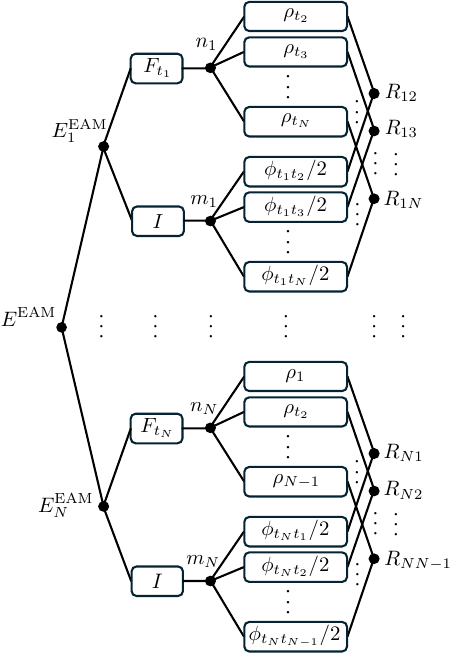}
    \caption{The schematic of EAM potential}
    \label{EAM}
\end{figure}

The total potential energy in the EAM is divided into two contributions, namely a pairwise part and a local density part: 
\begin{align}
    E^\mathrm{EAM}(\pmb{R}) & = \sum_{i=1}^{N} E_i^\mathrm{EAM}({\bm R}) \, , \\
    E_i^\mathrm{EAM}({\bm R}) & = F_{t_i} \! \left( n_i (\pmb{R}) \right) + \frac{1}{2} \sum_{j = 1, j\neq i}^N \phi_{t_i t_j}(R_{ij}) \, , \\
    n_i (\pmb{R}) &= \sum_{j=1, j \neq i}^N \rho_{t_j} (R_{ij}) \, .
\end{align}
The functions $F_{\alpha} (n)$, $\phi_{\alpha \beta}(R)$, and $\rho_{\alpha}(R)$ are constructed empirically or by fitting the potential energy calculated by DFT calculations. 
In actual simulations, the value of these functions is evaluated by linear interpolation. 
In terms of KAN, the local potential energy $E_i^{\rm EAM}$ can be rewritten as 
\begin{align}
E_i^\mathrm{EAM}(\pmb{R}) & = F_{t_i} \! \left(n_i  (\pmb{R}) \right) + I \left(m_i  (\pmb{R}) \right) \, , \\
I(x) & = x \, ,\\
m_i (\pmb{R}) & = \sum_{j=1, j \neq i}^N  \frac{1}{2} \phi_{t_i t_j}(R_{ij}) \, ,
\end{align}
where $t_i$ is the element type of the $i$-th atom.
The EAM potential can be represented by a type of KAN, as shown in Fig.~\ref{EAM}.
In terms of KAN, the EAM potential can be extended by adding more hidden layers and/or units.

\subsubsection{Aarificial neural network potential}
\begin{figure*}[t]
\includegraphics[width=0.6 \linewidth]{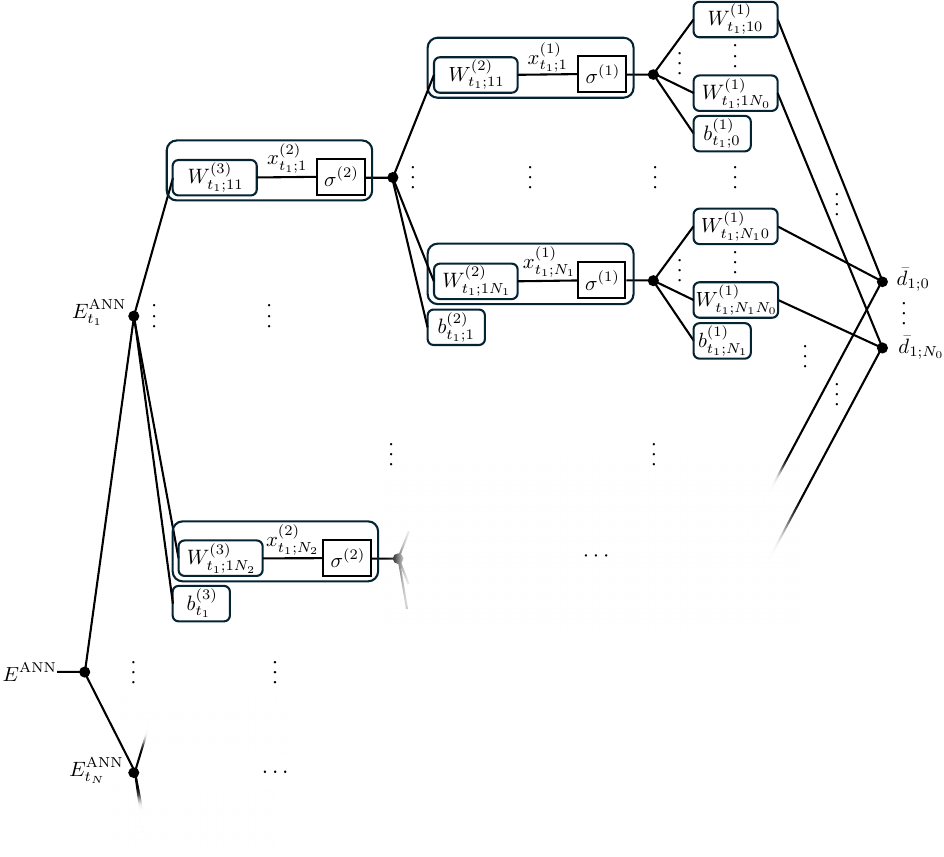}
\caption{The schematic of ANN potential. (a) The total energy is defined by the summation of atomic energies. (b) The atomic energy is given by the ANN, which is described by a KAN.}
\label{ANN}
\end{figure*}

There are various kinds of ANN potentials. 
In this paper, we focus on  Behler--Parrinello type neural networks \cite{Behler2007-kn}. 
In their formulation, the potential energy is assumed to be constructed from a sum of fictitious atomic energies, similar to the LJ and EAM potentials. 
Furthermore, the atomic energy is considered a functional of the atomic environment around the central atom, defined inside a sphere with a cutoff radius $R_\mathrm{c}$.
The feature values of the atomic environment, called descriptors, were introduced to ensure the translational and rotational invariance of the atomic energy.
Therefore, the total and atomic energies are given as
\begin{align}
    E^\mathrm{ANN} (\pmb{R}) & = \sum_{i=1}^N E_{t_i}^\mathrm{ANN} \left(\pmb{d}_i (\pmb{R}; {R_\mathrm{c}}) \right) ,
\end{align}
where $E^\mathrm{ANN}$, $E_{t_i}^\mathrm{ANN}$, and $\pmb{d}_i$ are the total energy, the atomic energy for the $i$-th atom, and descriptor of the atomic environment for the centered $i$-th atom, respectively.
There are many kinds of descriptors.
For example, in the case of the Chebyshev descriptor \cite{Artrith2017-ow}, the descriptor is determined as 
\begin{align}
    \pmb{d}_i (\pmb{R}; R_\mathrm{c}) & =
    \begin{pmatrix}
        d_{i;0} (\pmb{R}; R_\mathrm{c}) \\
        \vdots \\
        d_{i;N_0} (\pmb{R}; R_\mathrm{c})
    \end{pmatrix}
    \nonumber \\
    & =
    \begin{pmatrix}
    \pmb{d}_{i}^{(\mathrm{R})} (\pmb{R}; R_\mathrm{c}) \\
    \pmb{d}_{i}^{(\mathrm{A})} (\pmb{R}; R_\mathrm{c}) 
    \end{pmatrix} , \\
    \pmb{d}_{i}^{(\mathrm{R})} (\pmb{R}; R_\mathrm{c}) &=     
    \begin{pmatrix}
    \pmb{d}_{i}^{(\mathrm{r})} (\pmb{R}; R_\mathrm{c}) \\
    \pmb{d}_{i}^{(\mathrm{r,w})} (\pmb{R}; R_\mathrm{c}) 
    \end{pmatrix}, \nonumber \\
    &=\begin{pmatrix}
    (d_{i;0}^{(\mathrm{r})} (\pmb{R}; R_\mathrm{c}),\cdots, d_{i;N^{(\mathrm{r})}}^{(\mathrm{r})} (\pmb{R}; R_\mathrm{c}))^{\rm T} \\
    (d_{i;0}^{(\mathrm{r,w})} (\pmb{R}; R_\mathrm{c}),\cdots,d_{i;N^{(\mathrm{r})}}^{(\mathrm{r,w})} (\pmb{R}; R_\mathrm{c}) )^{\rm T}
    \end{pmatrix}, \\
    \pmb{d}_{i}^{(\mathrm{A})} (\pmb{R}; R_\mathrm{c}) &=    
    \begin{pmatrix}
    \pmb{d}_{i}^{(\mathrm{a})} (\pmb{R}; R_\mathrm{c}) \\
    \pmb{d}_{i}^{(\mathrm{a,w})} (\pmb{R}; R_\mathrm{c}) 
    \end{pmatrix} \nonumber \\
    &=\begin{pmatrix}
    (d_{i;0}^{(\mathrm{a})} (\pmb{R}; R_\mathrm{c}),\cdots, d_{i;N^{(\mathrm{a})}}^{(\mathrm{a})} (\pmb{R}; R_\mathrm{c}))^{\rm T} \\
    (d_{i;0}^{(\mathrm{a,w})} (\pmb{R}; R_\mathrm{c}),\cdots ,d_{i;N^{(\mathrm{a})}}^{(\mathrm{a,w})} (\pmb{R}; R_\mathrm{c}) )^{\rm T}
    \end{pmatrix}
\end{align}
    \begin{align}
    d_{i;s}^{(\mathrm{r})} (\pmb{R}; R_\mathrm{c}) & = \sum_{j=1, j \ne i}^N T_s ( \tilde{R} (\pmb{R}_{ij}; R_\mathrm{c}) )  f_\mathrm{c} (\pmb{R}_{ij}; R_\mathrm{c}) ,\\
    d_{i;s}^{(\mathrm{r}, \mathrm{w})} (\pmb{R}; R_\mathrm{c}) & = \sum_{j=1, j \ne i}^N T_s ( \tilde{R} (\pmb{R}_{ij}; R_\mathrm{c}) )  f_\mathrm{c} (\pmb{R}_{ij}; R_\mathrm{c}) w_{t_j}, \label{eq:c2} \\
    d_{i;s}^{(\mathrm{a})} (\pmb{R}; R_\mathrm{c})  &= \sum_{\substack{j,k=1, j\ne k\\ j \ne i, k \ne i}}^N T_s \! \left( c (\pmb{R}_{ij},\pmb{R}_{ik}) \right) \nonumber \\
    & \qquad \qquad \quad {} \times f_\mathrm{c} (\pmb{R}_{ij}; R_\mathrm{c}) f_\mathrm{c}(\pmb{R}_{ik}; R_\mathrm{c}) , \\
    d_{i;s}^{(\mathrm{a}, \mathrm{w})} (\pmb{R}; R_\mathrm{c})  &= \sum_{\substack{j,k=1, j\ne k\\ j \ne i, k \ne i}}^N T_s \! \left( c (\pmb{R}_{ij},\pmb{R}_{ik}) \right) \nonumber \\
    & \qquad \qquad {} \times f_\mathrm{c} (\pmb{R}_{ij}; R_\mathrm{c}) f_\mathrm{c}(\pmb{R}_{ik}; R_\mathrm{c}) w_{t_j}  w_{t_k}, \label{eq:c3}
\end{align}
where 
\begin{align}
    \bar{R}(\pmb{R}; R_\mathrm{c}) & =  2\frac{|\pmb{R}|}{R_\mathrm{c}}  - 1, \\
    c(\pmb{R},\pmb{R}') & = \frac{\pmb{R} \cdot \pmb{R}'}{|\pmb{R}| |\pmb{R}'|}, \\
    N_0 &= 2N^{(\mathrm{r})} + 2N^{(\mathrm{a})} + 3,
\end{align}
$T_s (x)$ is the $s$-th Chebyshev polynomial of the first kind, and $f_\mathrm{c}(\pmb{R}; R_\mathrm{c})$ is a cutoff function that smoothly goes to zero at $R_\mathrm{c}$.
For example, the following function is used for many cases:
\begin{equation}
    f_\mathrm{c} (\pmb{R}; R_\mathrm{c}) = 
    \begin{cases}
    \displaystyle \frac{1}{2} \left[ \cos \! \left( \pi \frac{|\pmb{R}|}{R_\mathrm{c}} \right) + 1 \right] & \displaystyle (0 \le |\pmb{R}| \le R_\mathrm{c}) \\
    \displaystyle 0 & \displaystyle (R_\mathrm{c} < |\pmb{R}|)
    \end{cases} .
\end{equation}
We introduce the shorthand notations as
\begin{align}
    \pmb{d}^{(\cdot)}_i & = \pmb{d}^{(\cdot)}_i (\pmb{R}, R_\mathrm{c}) \, , \\
    d^{(\cdot)}_{i; s} & = d^{(\cdot)}_{i; s} (\pmb{R}, R_\mathrm{c}) \, , \\
    \bar{R}_{ij} & = \bar{R}(\pmb{R}_{ij}; R_\mathrm{c}) \, , \\
    c_{ijk} & = c(\pmb{R}_{ij},\pmb{R}_{ik}) \, , \\
    \bar{f}_\mathrm{c} (R_{ij}) & = f_\mathrm{c} (\pmb{R}_{ij}; R_\mathrm{c}) \, .
\end{align}
When one considers MLP with two hidden layers, the $i$-th atomic energy is given as 
\begin{align}
    & E_{t_i}^\mathrm{ANN}(\pmb{d}_i ) = \sum_{\ell_{2}=1}^{N_2} W^{(3)}_{t_i; 1 \ell_{2}} x^{(2)}_{i; \ell_2} + b_{t_i}^{(3)} \, , \\
    & x^{(2)}_{i; \ell_2} = \sigma^{(2)} \! \! \left( \sum_{\ell_1=1}^{N_1} W^{(2)}_{t_i; \ell_{2} \ell_{1}} x^{(1)}_{i; \ell_1}+ b_{t_i; \ell_2}^{(2)} \right), \\
    & x^{(1)}_{i; \ell_1} = \sigma^{(1)} \! \! \left( \sum_{\ell_0=0}^{N_0} W^{(1)}_{t_i; \ell_{1} \ell_{0}} \bar{d}_{i; \ell_0} (\pmb{R}; {R_\mathrm{c}}) + b_{t_i; \ell_1}^{(1)} \right), \label{eq:x1}
\end{align}
where $\sigma^{(l)}(x)$ is an activation function of $l$-th layer such as $\tanh(x)$, and $\bar{\pmb{d}}_i$ is the normalized descriptor, of which element is defined as 
\begin{align}
    & \, \, \, \quad \bar{d}_{i;\ell_0} \nonumber \\
    & = \frac{d_{i; \ell_0} - m_{t_i; \ell_0}}{\sigma_{t_i; \ell_0}} \nonumber \\
    & =
    \begin{cases}
    \displaystyle \frac{d_{i;\ell_0}^{(\mathrm{R})} - m_{t_i; \ell_0}^{(\mathrm{R})}}{\sigma_{t_i; \ell_0}^{(\mathrm{R})}} & (\ell_0 \le 2N^{(\mathrm{r})}+1) \\
    \displaystyle \frac{d_{i;\ell_0 - 2N^{(\mathrm{r})} - 2}^{(\mathrm{A})} - m_{t_i; \ell_0 - 2N^{(\mathrm{r})} - 2}^{(\mathrm{A})}}{\sigma_{t_i; \ell_0 - 2N^{(\mathrm{r})} - 2}^{(\mathrm{A})}} & (2N^{(\mathrm{r})}+1 < \ell_0)
    \end{cases} ,
\end{align}
where
\begin{align}
    m_{t_i; \ell_0} & =
    \begin{cases}
    \displaystyle m_{t_i; \ell_0}^{(\mathrm{R})} & (\ell_0 \le 2N^{(\mathrm{r})}+1) \\
    \displaystyle m_{t_i; \ell_0 - 2N^{(\mathrm{r})} - 2}^{(\mathrm{A})} & (2N^{(\mathrm{r})} +1< \ell_0)
    \end{cases} , \\
    \sigma_{t_i; \ell_0} & =
    \begin{cases}
    \displaystyle \sigma_{t_i; \ell_0}^{(\mathrm{R})} & (\ell_0 \le 2N^{(\mathrm{r})}+1) \\
    \displaystyle \sigma_{t_i; \ell_0 - N2^{(\mathrm{r})} - 2}^{(\mathrm{A})} & (2N^{(\mathrm{r})} +1< \ell_0)
    \end{cases} ,
\end{align}
Here, $m_{i_0}$ and $\sigma_{i_0}$ are the mean value and standard deviation in a total data set. 
The ANN potential can be represented as a KAN, as shown in Fig.~\ref{ANN}.

Let us discuss the relation between the KAN and ANN potential. 
We focus on the first layer shown in Eq.~(\ref{eq:x1}). 
This equation is rewritten as 
\begin{align}
    x^{(1)}_{i; \ell_1} &= \sigma^{(1)} \! \left( z^\mathrm{(r)}_{i; \ell_1} + z^\mathrm{(a)}_{i; \ell_1} +  b_{t_i; \ell_1}^{(1)} \right), 
    \label{eq:xz3}
\end{align}
where we define $z^{(\mathrm{r})}_{i; \ell_1}$ and $z^{(\mathrm{a})}_{i; \ell_1}$ as 
\begin{align}
    z^{(\mathrm{r})}_{i; \ell_1} & = \sum_{\ell_0=0}^{2N^{(\mathrm{r})}+1} W^{(1)}_{t_i; \ell_{1} \ell_{0}}  \frac{d_{i; \ell_0}^{(\mathrm{R})} - m_{t_i; \ell_0}^{(\mathrm{R})}}{\sigma_{t_i; \ell_0}^{(\mathrm{R})}} \nonumber \\
    & = \sum_{\ell_0=0}^{2N^{(\mathrm{r})}+1} \bar{W}^{(1, \mathrm{R})}_{t_i; \ell_1 \ell_0} d_{i; \ell_0}^{(\mathrm{R})}+ b_{t_i; \ell_1}^{(\mathrm{R})} \, , \label{eq:zalpha}\\
    \bar{W}^{(1, \mathrm{R})}_{t_i; \ell_1 \ell_0} & = \frac{W^{(1)}_{t_i; \ell_{1} \ell_{0}} }{\sigma_{t_i; \ell_0}^{(\mathrm{R})}} \, ,\\
    b_{t_i; \ell_1}^{(\mathrm{R})} & = -\sum_{\ell_0=0}^{N^{(\mathrm{r})}} W^{(1)}_{t_i; \ell_{1} \ell_{0}}  \frac{m_{t_i; \ell_0}^{(\mathrm{R})}}{\sigma_{t_i; \ell_0}^{(\mathrm{R})}} \, ,
\end{align}
and
\begin{align}
    z^{(\mathrm{a})}_{i; \ell_1} & = \sum_{\ell_0'=0}^{2N^{(\mathrm{a})}+1} W^{(1)}_{t_i; \ell_{1} \ell_{0}' + 2N^{(\mathrm{r})} + 2} \frac{d_{i; \ell_0'}^{(\mathrm{A})} - m_{t_i; \ell_0'}^{(\mathrm{A})}}{\sigma_{t_i; \ell_0'}^{(\mathrm{A})}} \nonumber \\
    & = \sum_{\ell_0'=0}^{2N^{(\mathrm{a})}+1} \bar{W}^{(1, \mathrm{A})}_{t_i; \ell_{1} \ell_{0}'} d_{i; \ell_0'}^{(\mathrm{A})}+ b_{t_i; \ell_1}^{(\mathrm{A})} \, , \label{eq:zalphaa}\\
    \bar{W}^{(1, \mathrm{A})}_{t_i; \ell_1 \ell_0'} & = \frac{W^{(1)}_{t_i; \ell_{1} \ell_{0}' + 2N^{(\mathrm{r})} + 2} }{\sigma_{t_i; \ell_0'}^{(\mathrm{A})}} \, , \\
    b_{t_i; \ell_1}^{(\mathrm{A})} & = -\sum_{\ell_0'=0}^{2N^{(\mathrm{a})}+1} W^{(1)}_{t_i; \ell_{1} \ell_{0}' + 2N^{(\mathrm{r})} + 2} \frac{m_{t_i; \ell_0'}^{(\mathrm{A})}}{\sigma_{t_i; \ell_0'}^{(\mathrm{A})}} \, ,
\end{align}
respectively.
By substituting Eqs.~(\ref{eq:c2}) and (\ref{eq:c3}) into Eq.~(\ref{eq:zalpha}), $z^{(\mathrm{r})}_{i_1}$ and  $z^{(\mathrm{a})}_{i_1}$ are given as 
\begin{align}
    z^\mathrm{(r)}_{i; \ell_1} &= \sum_{j=1, j\ne i}^{N} \Phi_{t_i,t_j; \ell_1}^\mathrm{(r)} (\bar{R}_{ij} ) \, \bar{f}_\mathrm{c} (R_{ij}) + b_{t_i; \ell_1}^\mathrm{(R)}, \\
    z^\mathrm{(a)}_{\ell_1} &= \sum_{\substack{j,k=1, j\ne k\\ j \ne i, k \ne i}}^N \Phi_{t_i,t_j,t_k; \ell_1}^\mathrm{(a)} (c_{ijk}) \bar{f}_\mathrm{c} (R_{ij}) \bar{f}_\mathrm{c} (R_{ik}) + b_{t_i; \ell_1}^\mathrm{(A)},
\end{align}
where   
\begin{align}
    \Phi_{t_i,t_j; \ell_1}^\mathrm{(r)} (R) & = \phi_{t_i; \ell_1}^\mathrm{(r)}(R_{ij} ) + \phi_{t_i; \ell_1}^\mathrm{(r,w)}(R_{ij} ) w_{t_j},\\
    \Phi_{t_i,t_j,t_k; \ell_1}^\mathrm{(a)} (\cos \theta) &=  \phi_{t_i; \ell_1}^\mathrm{(a)}(\cos \theta) +  \phi_{t_i; \ell_1}^\mathrm{(a,w)}(\cos \theta) w_{t_j}w_{t_k}, \\
    \phi_{t_i; \ell_1}^\mathrm{(r)}(R) &=  \sum_{\ell_0=0}^{N^\mathrm{(r)}} \bar{W}^{(1,\mathrm{R})}_{t_i; \ell_{1} \ell_0} T_{\ell_0} (R) , \label{eq:phi1} \\
    \phi_{t_i; \ell_1}^\mathrm{(r,w)}(R) &=  \sum_{\ell_0=0}^{N^\mathrm{(r)}} \bar{W}^{(1,\mathrm{R})}_{t_i; \ell_{1} \ell_0+N^\mathrm{(r)}+1} T_{\ell_0} (R) , \label{eq:phi1w} \\
    \phi_{t_i; \ell_1}^\mathrm{(a)}(\cos \theta) &=  \sum_{\ell_0'=0}^{N^\mathrm{(a)}} \bar{W}^{(1,\mathrm{A})}_{t_i; \ell_{1} \ell_0'} T_{\ell_0'}(\cos \theta),  \label{eq:phi2} \\
     \phi_{t_i; \ell_1}^\mathrm{(a,w)}(\cos \theta) &=  \sum_{\ell_0'=0}^{N^\mathrm{(a)}} \bar{W}^{(1,\mathrm{A})}_{t_i; \ell_{1} \ell_0'+N^\mathrm{(a)}+1} T_{\ell_0'}(\cos \theta). \label{eq:phi2w}
\end{align}
In terms of KAN, the one-dimensional non-linear functions $\phi_{t_i; \ell_1}^\mathrm{(r)} (R)$, $\phi_{t_i; \ell_1}^\mathrm{(r,w)} (R_{ij})$, $\phi_{t_i; \ell_1}^\mathrm{(a)}(\cos \theta)$, and $\phi_{t_i; \ell_1}^\mathrm{(a,w)}(\cos \theta)$ are regarded as activation functions in the KAN structure. 
The above equations show that these non-linear functions are expanded by the Chebyshev polynomials.

\section{Kolmogorov--Arnold Network descriptor}
\subsection{Definition}
We propose a trainable descriptor called ``KAN descriptor'' for interatomic potentials, defined as 
\begin{equation}
    \tilde{\pmb{z}}_i (\pmb{R}; R_\mathrm{c})  =
    \begin{pmatrix}
        \tilde{z}_{i; 1} (\pmb{R}; R_\mathrm{c})\\
        \tilde{z}_{i; 2} (\pmb{R}; R_\mathrm{c})\\
        \vdots \\
        \tilde{z}_{i: \tilde{N}_0} (\pmb{R}; R_\mathrm{c})
    \end{pmatrix},
\end{equation}
where
\begin{align}
    & \quad \, \, \tilde{z}_{i; \ell_0} (\pmb{R}; R_\mathrm{c}) \nonumber \\
    &= \sum_{j=1, j \ne i}^N \tilde{\Psi}^{(\mathrm{r})}_{t_i t_{j}; \ell_0}(\pmb{R}_{ij};R_{\rm c}) \nonumber \\
    & \quad \, {} +  \sum_{\substack{j,k=1, j\ne k\\ j \ne i, k \ne i}}^N \tilde{\Psi}^{(\mathrm{a})}_{t_i t_{j}t_{k}; \ell_0}(\pmb{R}_{ij},\pmb{R}_{ik};R_{\rm c}) + \tilde{b}_{t_i;\ell_0} \, , \\
    &= \sum_{j=1, j \ne i}^N \tilde{\Phi}^{(\mathrm{r})}_{t_i t_{j}; \ell_0}(\pmb{R}_{ij}) f_{\rm c}(\pmb{R}_{ij};R_{\rm c}) \nonumber \\
    & \quad \, {} +  \sum_{\substack{j,k=1, j\ne k\\ j \ne i, k \ne i}}^N \tilde{\Phi}^{(\mathrm{a})}_{t_i t_{j}t_{k}; \ell_0}(\pmb{R}_{ij},\pmb{R}_{ik}) f_{\rm c}(\pmb{R}_{ij};R_{\rm c})  f_{\rm c}(\pmb{R}_{ik};R_{\rm c}) \nonumber \\
    & \quad \, {}  + \tilde{b}_{t_i;\ell_0}.
\end{align}
Here, $\tilde{\Phi}^{(\mathrm{r})}_{t_it_{j}; \ell_0}(\pmb{R}_{ij})$ and $\tilde{\Phi}^{(\mathrm{a})}_{t_it_{j}t_{k}; \ell_0}(\pmb{R}_{ij},\pmb{R}_{ik})$ are radial (two-body) and angular (three-body) non-linear functions, which should be optimized. 

\begin{figure}[th]
    \centering
    \includegraphics[width=0.8 \linewidth]{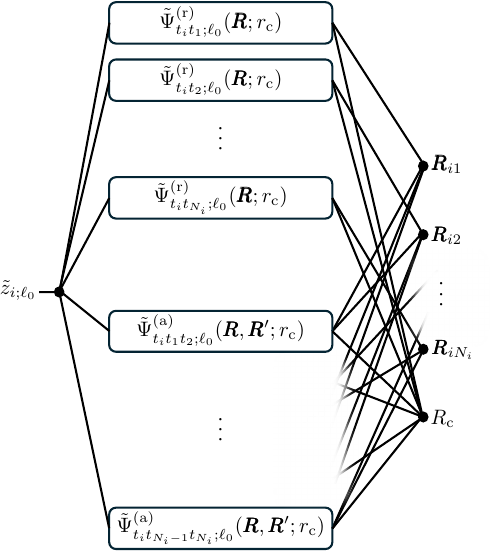}
    \caption{The schematic of $i_0$-th element of the Kolmogorov--Arnold network descriptor.}
    \label{KAN_descriptor}
\end{figure}

\begin{figure*}[t]
\includegraphics[width=0.6 \linewidth]{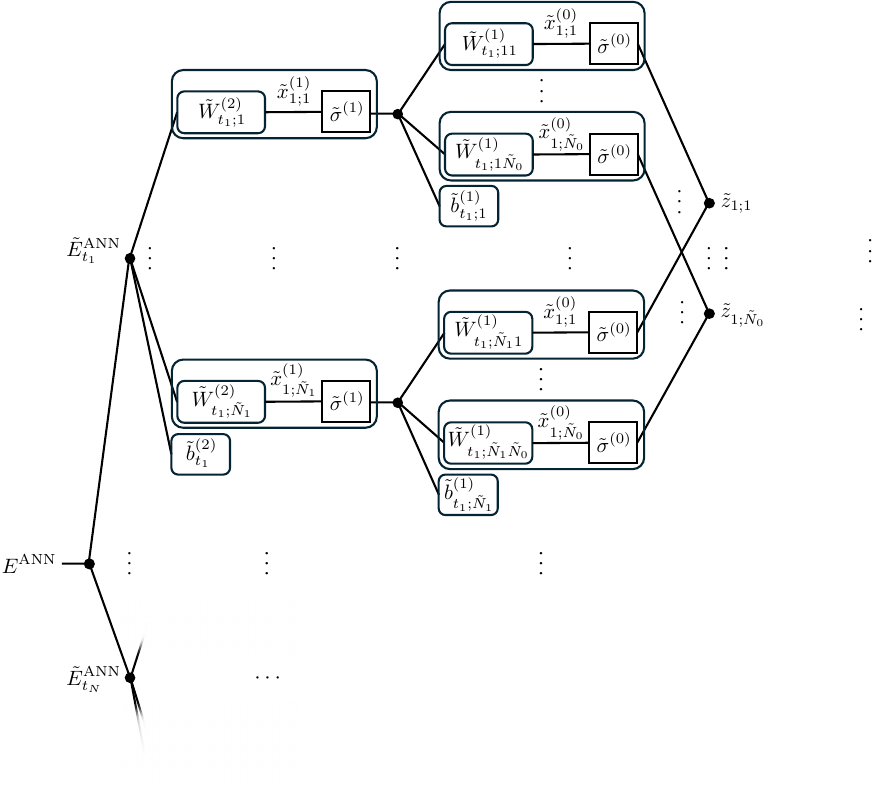}
\caption{The schematic of artificial neural networks potential with the Kolmogorov--Arnold network descriptors.}
\label{ANN_KAN}
\end{figure*}

The total energy is expressed as
\begin{align}
    E^\mathrm{ANN}(\pmb{R}) & = \sum_{i=1}^N \tilde{E}^\mathrm{ANN}_{t_i} (\tilde{\pmb{z}}_i (\pmb{R}; R_\mathrm{c})) , 
\end{align}
where $\tilde{E}^\mathrm{ANN}_{t_i} (\tilde{\pmb{z}}_i (\pmb{R}; R_\mathrm{c}))$ is the $i$-th atomic potential constructed by neural networks with the KAN descriptor $\tilde{\pmb{z}}_i (\pmb{R}; R_\mathrm{c}) $ as shown in Fig.~\ref{KAN_descriptor}.  
It should be noted that both the function $\tilde{E}^\mathrm{ANN}_{t_i}$ and the KAN descriptor $\tilde{\pmb{z}}_i (\pmb{R}; R_\mathrm{c})$ are trainable. 
When one considers MLP with one hidden layer, the $i$-th atomic energy is expressed as 
\begin{align}
    \tilde{E}^\mathrm{ANN}_{t_i} (\tilde{\pmb{z}}_i (\pmb{R}; R_\mathrm{c})) & = \sum_{\ell_1=1}^{\tilde{N}_1} \tilde{W}^{(2)}_{t_i; \ell_1} \tilde{x}^{(1)}_{i; \ell_1} + \tilde{b}^{(2)}_{t_i} \, , \\
    \tilde{x}^{(1)}_{i; \ell_1} & = \tilde{\sigma}^{(1)} \! \left( \sum_{\ell_0=1}^{\tilde{N}_0} \tilde{W}^{(1)}_{t_i; \ell_1 \ell_0} \tilde{x}^{(0)}_{i; \ell_0} + \tilde{b}^{(1)}_{t_i; \ell_1} \right) , \\
    \tilde{x}^{(0)}_{i; \ell_0} & = \tilde{\sigma}^{(0)} \! \left( \tilde{z}_{i;\ell_0} (\pmb{R}; R_\mathrm{c}) \right).
\end{align}
The network architecture with KAN descriptors is shown in Fig.~\ref{ANN_KAN}.

In the original KAN layer shown in Eq.~(\ref{eq:KANlayer}), non-linear functions $\phi_{l,i_l,i_{l-1}}(x)$ depend on indices of input and output neurons. 
However, in the KAN descriptor, the non-linear functions $\tilde{\Phi}^{(\mathrm{r})}_{t_i t_{j}; \ell_0}(\pmb{R}_{ij})$ and $\tilde{\Phi}^{(\mathrm{a})}_{t_i t_{j}t_{k}; \ell_0}(\pmb{R}_{ij},\pmb{R}_{ik})$ do not depend on an index of input neurons and only on the kinds of atoms and an index of output neurons since atoms with the same types are indistinguishable.
In Fig.~\ref{KAN_descriptor}, it looks there are $N_i$ distinct functions $\tilde{\Phi}^\mathrm{(r)}_{t_i t_j; \ell_0}$ and $N_i^2$ distinct functions $\tilde{\Phi}^{\rm (a)}_{t_i t_j t_k; \ell_0}$ for $\ell_0$. 
However, the number of non-linear functions of the KAN descriptor is usually much smaller than $N_i$ or $N_i^2$ and depends on the number of the kind of atoms. 
For example, consider 10 water molecules surrounding the $i$-th hydrogen atom in an H$_2$O system.
In this case, only two $\tilde{\Phi}^\mathrm{(r)}$ functions ($\tilde{\Phi}^\mathrm{(r)}_{\mathrm{H H}; \ell_0}$ and $\tilde{\Phi}^\mathrm{(r)}_{\mathrm{H O}; \ell_0}$) and three $\tilde{\Phi}^\mathrm{(a)}$ functions ($\tilde{\Phi}^\mathrm{(r)}_{\mathrm{HHH};\ell_0}$, $\tilde{\Phi}^\mathrm{(r)}_{\mathrm{HHO};\ell_0}$, and $\tilde{\Phi}^\mathrm{(r)}_{\mathrm{HOO};\ell_0}$) are required although $N_i = 30$.

These non-linear functions should be optimized to approximate DFT energy. 
The non-linear functions can be expanded using B-spline functions, Gaussian radial distribution functions, or Chebyshev polynomials. 
During the training procedure, the coefficients of the non-linear functions are adjusted. 
Once accurate non-linear functions with appropriate coefficients are obtained, values can be evaluated using linear or spline interpolation. 
In this study, we employ the Chebyshev polynomials as
\begin{align}
    \tilde{\Phi}^\mathrm{(r)}_{t_i t_j; \ell_0}(\pmb{R}_{ij}) &= \tilde{\varphi}^\mathrm{(r,0)}_{t_i; \ell_0} (\bar{R}_{ij})  + \tilde{\varphi}^\mathrm{(r,1)}_{t_i; \ell_0} (\bar{R}_{ij}) w_{t_j}, \\   
    \tilde{\varphi}^{(\mathrm{r},n)}_{t_i; \ell_0} (R)  &= \sum_{s=0}^{\tilde{N}^\mathrm{(r)}} \tilde{W}^{(\mathrm{r},n)}_{t_i; \ell_0 s} \, T_s (R),\\
    \tilde{\Phi}^{(\mathrm{a})}_{t_i t_{j}t_{k}; \ell_0}(\pmb{R}_{ij},\pmb{R}_{ik})  &= \tilde{\varphi}^\mathrm{(a,0)}_{t_i; \ell_0} (c_{ijk}) + 
     \tilde{\varphi}^\mathrm{(a,0)}_{t_i; \ell_0} (c_{ijk}) w_{t_j} w_{t_k},\\
     \tilde{\varphi}^{(\mathrm{a},n)}_{t_i; \ell_0} (\cos \theta) & = \sum_{s=0}^{\tilde{N}^\mathrm{(a)}} \tilde{W}^{(\mathrm{a},n)}_{t_i; \ell_0 s} \, T_s (\cos \theta).
\end{align}
We mention the representation ability of the non-linear functions.
The number of the trainable parameters is $2 \tilde{N}_0 (\tilde{N}^\mathrm{(r)}+1) + 2 \tilde{N}_0 (\tilde{N}^\mathrm{(a)} + 1)$ for the KAN descriptor $\tilde{\pmb{z}}_i (\pmb{R}; R_\mathrm{c})$. 
In addition, these non-linear functions depend on the kinds of atoms as same as the original Chebyshev descriptor \cite{Artrith2017-ow}.
Therefore, the KAN descriptor can avoid dimensional explosion with respect to the number of atom kinds like the symmetry functions \cite{Behler2007-kn}.

\subsection{Relation between the Chebyshev and the Kolmogorov--Arnold network descriptors}
The relations between the coefficients of the non-linear functions of the KAN descriptor and the coefficients defined in Eqs.~(\ref{eq:xz3})--(\ref{eq:zalpha}) are expressed as 
\begin{align}
    \tilde{b}_{t_i;\ell_0} &=b_{t_i; \ell_0}^{(1)} + b_{t_i; \ell_0}^{(\mathrm{R})}  + b_{t_i; \ell_0}^{(\mathrm{A})}, \\
\tilde{W}^{(\mathrm{r},0)}_{t_i; \ell_0 n} &= \bar{W}^{(1, \mathrm{R})}_{t_i; \ell_0 n},  \\
\tilde{W}^{(\mathrm{r},1)}_{t_i; \ell_0 n} &= \bar{W}^{(1, \mathrm{R})}_{t_i; \ell_0 n+\tilde{N}^{(r)}+1}, \\
\tilde{W}^{(\mathrm{a},0)}_{t_i; \ell_0 n} &= \bar{W}^{(1, \mathrm{A})}_{t_i; \ell_0 n},  \\
\tilde{W}^{(\mathrm{a},1)}_{t_i; \ell_0 n} &= \bar{W}^{(1, \mathrm{A})}_{t_i; \ell_0 n+\tilde{N}^{(a)}+1}. 
\end{align}
In the previous Chebyshev descriptor, $\sigma_{t_i; \ell_0}^{(\mathrm{R})}$, $\sigma_{t_i; \ell_0}^{(\mathrm{A})}$, $m_{t_i; \ell_0}^{(\mathrm{R})}$ and $m_{t_i; \ell_0}^{(\mathrm{A})}$ are  calculated by a data set. 
In the KAN descriptor, these variables can be trainable. 
We note that the MLP with two-hidden layers with the Chebyshev descriptor can be regarded as the MLP with one-hidden layer with the KAN descriptor, since the KAN descriptor is trainable.  

\section{Results}
\subsection{Target system and method}
We consider iron with vacancies and dislocations. We use the training dataset available on GitHub as described in Ref.~\cite{Mori2023-yw}. The KAN descriptor and MLP are implemented using the machine-learning package \textit{Flux.jl} version 0.14.11, written in the Julia language\cite{Flux.jl-2018,innes:2018}. 
The sizes of the total dataset, training dataset, and test dataset are 6348, 5713, and 635, respectively. The energy of each atomic structure is calculated using {\it Quantum Espresso}, a first-principles calculation tool based on density functional theory \cite{Giannozzi2009-vk}. 
The details of the calculation are provided in Ref.~\cite{Mori2023-yw}. 
We adopt an MLP with one hidden layer for $\tilde{E}_i^\mathrm{ANN}(\tilde{\pmb{z}}_i(\pmb{R};R_\mathrm{c}))$. 
The dimension of the input vector $\tilde{\pmb{z}}$ and the number of units in the hidden layer were set as $\tilde{N}_0 = 10$ and $\tilde{N}_1 = 10$, respectively.
To train the ANN potential, we use the limited-memory Broyden--Fletcher--Goldfarb--Shanno (BFGS) algorithm implemented in the \textit{Optim.jl} package\cite{mogensen2018optim}. 
For each optimization with the limited-memory BFGS, we perform 10 iterations to reduce the loss function defined as
\begin{align}
{\rm MSE} &= \frac{1}{N_{\rm batch}} \sum_{k=1}^{N_{\rm batch}} \left( E^{\mathrm{FP}(k)} - E^{{\rm ANN} (k)} \right)^2.
\end{align}
Here, the batch size $N_{\rm batch}$ in the \textit{Optim.jl} package is 100. $E^{\mathrm{FP} (k)}$ and $E^{\rm ANN}(k)$ are the potential energies calculated by first-principles calculation and ANN, respectively.

\subsection{Shape of the non-linear functions in KAN descriptors}
\begin{figure}[t]
  \begin{minipage}[b]{1\linewidth}
    \includegraphics[width=\linewidth]{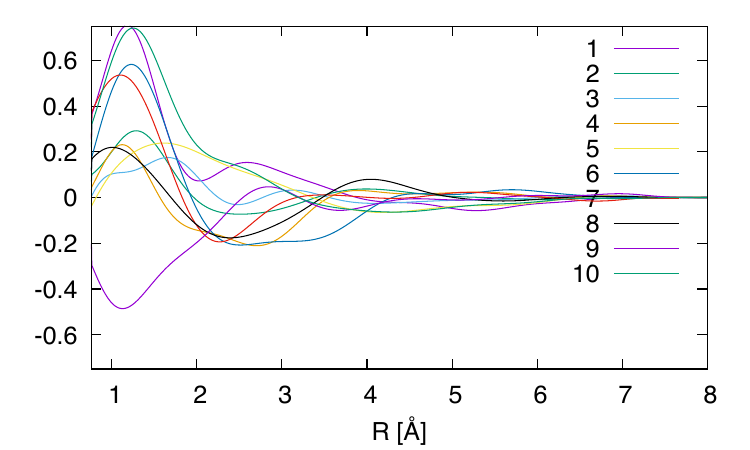}
  \end{minipage}
  \begin{minipage}[b]{1\linewidth}
    \includegraphics[width=\linewidth]{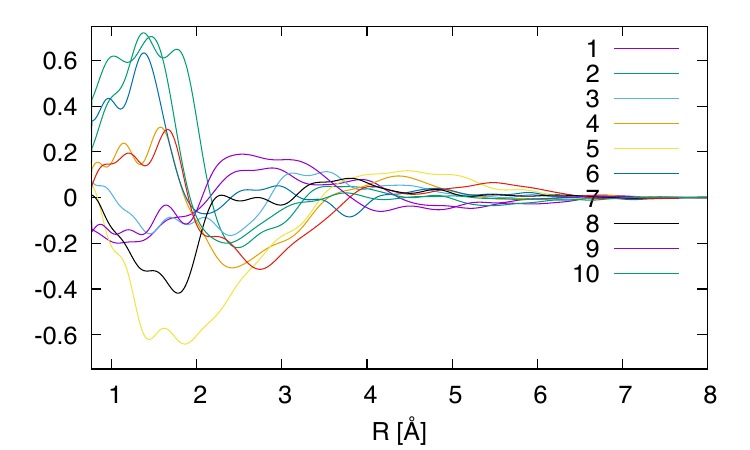}
  \end{minipage}
  \caption{Two-body non-linear functions $\tilde{\Phi}^{(\mathrm{r})}_{\mathrm{FeFe}; \ell_0}(\pmb{R}_{ij}) f_{\rm c}(\pmb{R}_{ij};R_{\rm c})$ in KAN descriptor. (Top panel)($\tilde{N}^\mathrm{(r)},\tilde{N}^\mathrm{(a)}) = (20,20)$ and  (Bottom panel) ($\tilde{N}^\mathrm{(r)},\tilde{N}^\mathrm{(a)}) = (50,20)$.}
  \label{fig:KAND}
\end{figure}

\begin{figure}
    \centering
    \includegraphics[width=\linewidth]{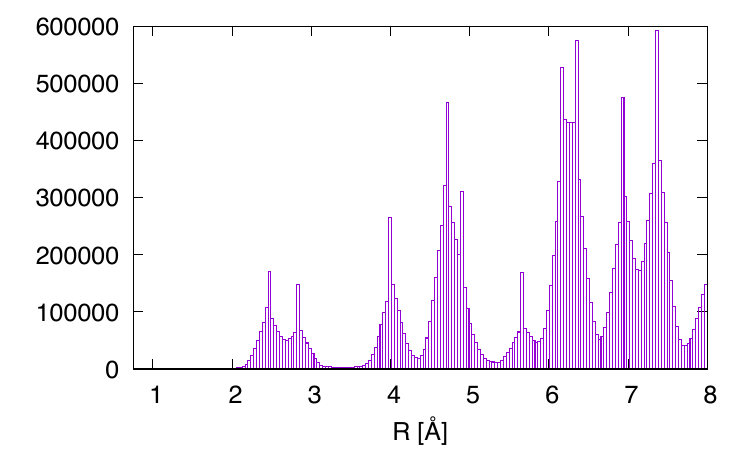}
    \caption{Radial distribution function in total data set}
    \label{fig:histogram}
\end{figure}

Let us discuss the shape of the non-linear functions in KAN descriptors. 
We consider two parameter sets ($\tilde{N}^\mathrm{(r)},\tilde{N}^\mathrm{(a)}) = (20,20)$ and $(50,20)$. 
As shown in Fig.~\ref{fig:KAND}, the shape of the non-linear functions in KAN descriptors $\tilde{\Phi}^{(\mathrm{r})}_{\mathrm{FeFe}; \ell_0}(\pmb{R}_{ij}) f_{\rm c}(\pmb{R}_{ij};R_{\rm c})$ depends on $\tilde{N}^\mathrm{(r)}$. 
In the region $R_{ij} < 2$ [\AA], the curves strongly depend on $\tilde{N}^\mathrm{(r)}$. 
It should be noted that there is no data with $R_{ij} < 2$ [\AA] in the training and test data set.
As shown in Fig.~\ref{fig:histogram}, the distribution of inter-atomic distance $R_{ij}$ is not uniform. 
In the region where no data is found in the training data set, the shape of the non-linear function in the KAN descriptor can not affect the quality of the ANN potential. 

\begin{figure}
    \centering
    \includegraphics[width=\linewidth]{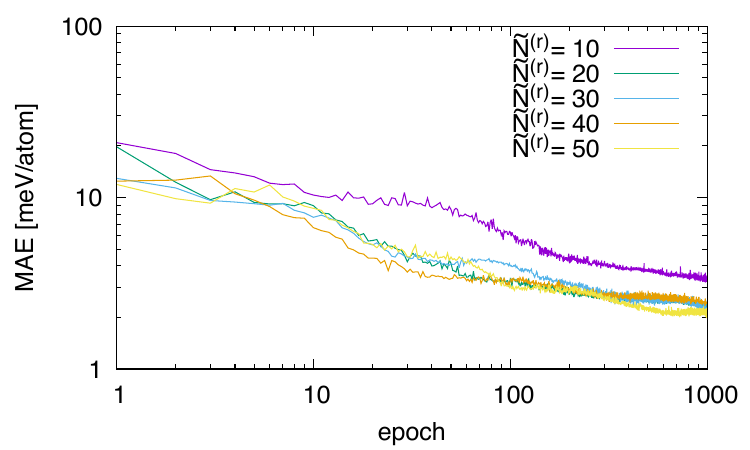}
    \caption{$\tilde{N}^\mathrm{(r)}$ dependence of the Mean absolute error. We fix the number of the angular Chebyshev basis ($\tilde{N}^\mathrm{(a)} = 20$), the dimension of the input vector constructed by the KAN descriptors and the number of the units of the hidden layer in the neural networks.  }
    \label{fig:lossdep}
\end{figure}

In terms of KANs, if the non-linear functions are well optimized with a fixed number of functions, the loss converges with increasing the number of the Chebyshev polynomials. 
As shown in Fig.~\ref{fig:lossdep}, the limits of the expressive power of the KAN descriptors are reached with increasing the number of the Chebyshev polynomials for radial non-linear functions.   

\subsection{Fast evaluation of the Kolmogorov--Arnold network descriptors}
We propose a method of a fast evaluation of the ANN potential with KAN descriptors. 
The conventional ANN potential constructed by the Chebyshev polynomial descriptor can be regarded as the ANN potential with KAN descriptors.
The computational cost to evaluate the KAN descriptor is ${\cal O}(\tilde{N}^\mathrm{(r)}) + {\cal O}(\tilde{N}^\mathrm{(a)}) $. 
With the use of linear or spline interpolation with $N$ points, the computational cost becomes ${\cal O}(\log N)$ which does not depend on the order of Chebyshev polynomials. 
We note that the computational cost of the forces can also be reduced with the use of interpolations.

\begin{figure}[tbp]
  \begin{minipage}[b]{1\linewidth}
    \includegraphics[width=\linewidth]{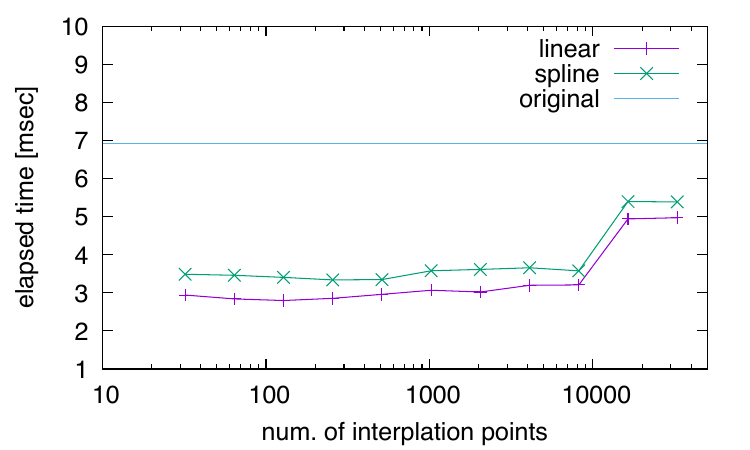}
  \end{minipage}
  \begin{minipage}[b]{1\linewidth}
    \includegraphics[width=\linewidth]{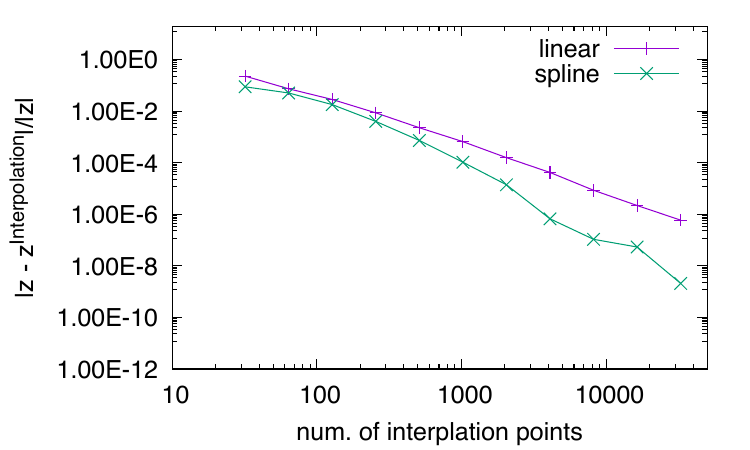}
  \end{minipage}
    \begin{minipage}[b]{1\linewidth}
    \includegraphics[width=\linewidth]{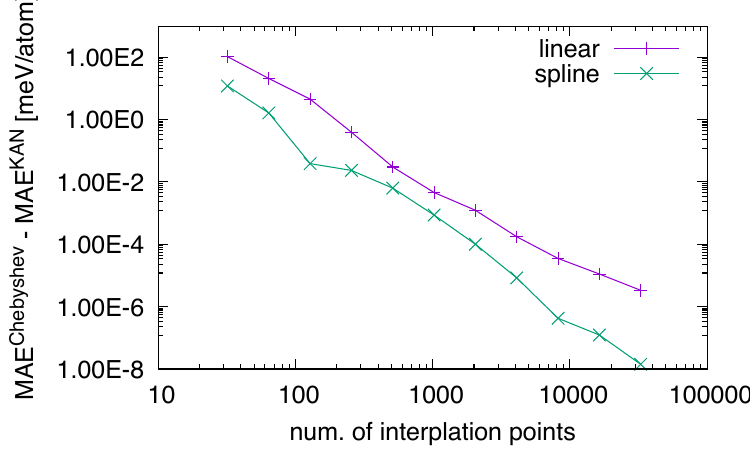}
  \end{minipage}
  \caption{(Top panel) Elapsed time to calculate the mean absolute error of the test data. (Middle panel) The residual norm of the KAN descriptors. (Bottom panel) The difference between the MAE of Chebyshev descriptors and that of KAN descriptors. The parameters are ($\tilde{N}^\mathrm{(r)},\tilde{N}^\mathrm{(a)}) = (50,20)$.}
  \label{fig:time}
\end{figure}

We discuss the interpolation point dependence of the computational time and accuracy. 
In the top panel in Fig.~\ref{fig:time}, we show the residual norm of the KAN descriptors defined as 
\begin{align}
    \frac{\left\|\tilde{\pmb{z}} - \tilde{\pmb{z}}^\mathrm{interpolation}\right\|}{\|\tilde{\pmb{z}}\|},
\end{align}
where $\tilde{\pmb{z}}^{\rm interpolation}$ is the input vector constructed by interpolations, and $\| \pmb{A} \| = \pmb{A}^2$. 
We adopt a linear interpolation and natural cubic spline interpolation implemented in \textit{Interpolation.jl}. 
The elapsed time is measured in MacBook Pro (14 inch, 2023) with Apple M2 Max processor. 
The parameters of the Chebyshev polynomial is $(\tilde{N}^\mathrm{(r)},\tilde{N}^\mathrm{(a)}) = (50,20)$. 
We should note that the code that we made written in Julia language might not be fully optimized. 
Therefore, we can discuss the elapsed time only qualitatively. 
As shown in Fig.~\ref{fig:time}, the elapsed time of each interpolation weakly depends on $\log N$, which is consistent with a theoretical value ${\cal O}(\log N)$.  
We also discuss the mean absolute loss determined as 
\begin{align}
    {\rm MAE} &= \frac{1}{M} \sum_{k=1}^M \left| E^{(k)} - E^{{\rm ANN} (k)} \right| \! ,
\end{align}
where $M$ is the number of the test data set. 
To discuss the accuracy of the interpolated KAN descriptors, we calculate the difference between the MAE with the ANN potential with Chebyshev polynomials and that with the interpolated KAN descriptors.
The MAE of the Chebyshev descriptors is $2.05796$ [meV/atom]. 
As shown in the bottom panel in Fig.~\ref{fig:time}, the difference reduces with increasing interpolation points.

We should note that, for actual machine-learning MD simulations, it is more better to implement the KAN descriptor to an open-source ANN potential package such as {\ae}net \cite{Artrith2017-ow}. 
Since the implementation might not be complicated, we will propose the software in the future.

\section{Conclusion}
In this study, we have integrated KAN into MD simulations to enhance the efficiency of potential energy models. 
Our investigation reveals that the KAN framework can represent commonly used potentials, such as the LJ, EAM, and ANN potentials. 
This reinterpretation allows for the application of KAN’s non-linear functions to introduce the KAN descriptor for ANN potentials. 
Our results demonstrate that by employing linear or cubic spline interpolations for these KAN functions, computational savings can be achieved without reducing accuracy. 
Future work will focus on further refining the KAN descriptors and integrating them into open-source ANN potential packages such as {\ae}net.

\begin{acknowledgments}
This work was partially supported by JSPS KAKENHI Grants Nos. 22H05111, 22H05114, 22K03539, and 23H03996.
This work was also partially supported by ``Joint Usage/Research Center for Interdisciplinary Large-scale Information Infrastructures (JHPCN)'' and ``High Performance Computing Infrastructure (HPCI)'' in Japan (Project ID: jh240028 and jh240059).
We would like to thank Dr. Mitsuhiro Itakura for the variable discussions.
Additionally, we are grateful to Dr. Hideki Mori for providing information about his training dataset for iron with vacancies and dislocations.
\end{acknowledgments}

\bibliography{kanpaper}

\end{document}